\documentclass[10pt,conference]{IEEEtran}
\IEEEoverridecommandlockouts

\usepackage{cite}
\usepackage{amsmath,amssymb,amsfonts}
\usepackage{algorithm}
\usepackage{algpseudocode}
\usepackage{graphicx}
\usepackage{textcomp}
\usepackage{xcolor}
\usepackage{subcaption}
\usepackage{arydshln}
\usepackage{multirow}
\def\BibTeX{{\rm B\kern-.05em{\sc i\kern-.025em b}\kern-.08em
    T\kern-.1667em\lower.7ex\hbox{E}\kern-.125emX}}

\begin{document}
\title{Joint Ex-Post Location Calibration and Radio Map Construction under Biased Positioning Errors
\thanks{This work was supported by JST, PRESTO under Grant Number JPMJPR23P3, and JST, CRONOS, Grant Number JPMJCS24N1.}
}

\author{\IEEEauthorblockN{Koki~Kanzaki~and~Koya~Sato}
\IEEEauthorblockA{Artificial Intelligence eXploration Research Center,\\
The University of Electro-Communications, Chofu, Tokyo, Japan\\
Email: k-kanzaki@uec.ac.jp, k\_sato@ieee.org
}
}

\maketitle

\begin{abstract}
This paper proposes a high-accuracy radio map construction method tailored for environments where location information is affected by bursty errors.
Radio maps are an effective tool for visualizing wireless environments.
Although extensive research has been conducted on accurate radio map construction, most existing approaches assume noise-free location information during sensing.
In practice, however, positioning errors ranging from a few to several tens of meters can arise due to device-based positioning systems (e.g., GNSS). Ignoring such errors during inference can lead to significant degradation in radio map accuracy.
This study highlights that these errors often tend to be biased when using mobile devices as sensors. We introduce a novel framework that models these errors together with spatial correlation in radio propagation by embedding them as tunable parameters in the marginal log-likelihood function. This enables ex-post calibration of location uncertainty during radio map construction.
Numerical results based on practical human mobility data demonstrate that the proposed method can limit RMSE degradation to approximately 0.25–0.29 dB, compared with Gaussian process regression using noise-free location data, whereas baseline methods suffer performance losses exceeding 1 dB.
\end{abstract}


\section{Introduction}
\label{sect:introduction}
Large-scale Internet-of-Things (IoT) systems have attracted significant interest in recent years.
For example, practical assessments and implementations of IEEE 802.11ah, a wireless LAN standard capable of providing communication coverage of up to approximately 1 km, are actively progressing, with expectations of a broadening scope of applications\cite{10531711}.
Typically, IoT systems are required to work with high reliability and efficiency.
Achieving these requirements depends on accurate prediction of radio propagation characteristics over extensive geographical areas; however, such prediction is very challenging owing to path loss, shadowing and multi-path fading.
To deal with this problem, the concept of a radio map\,\cite{zeng_comst2024} has recently gained substantial attention as a tool for high-accuracy wireless environment predictions.
\par
A radio map visualizes spatial variation of the received signal power values.
It can provide a statistical information on the wireless channel, thereby enhancing performance of wireless applications, such as wireless resource optimization\cite{bi-wirelesscommun2019}.
Recent works have revealed crowdsourcing-based approaches can accurately construct radio maps\cite{duCRCLocCrowdsourcingBasedRadio2022}.
They leverage sensing capabilities in user devices, such as smartphones and autonomous vehicles, to collect and upload radio information with their sensing location information to a cloud server.
The server then estimates spatial variations in received signal power values using spatial interpolation techniques\cite{xuRadioEnvironmentMap2021, romeroRadioMapEstimation2022}.
\par
Most of these methods assume the noiseless position information in sensing.
However, these positions are obtained via the device's positioning functions, which may inherently contain errors; for instance, a global navigation satellite system (GNSS) can incur errors ranging from several meters to several tens of meters, suggesting that regression analysis without accounting for these errors risks degrading the accuracy of the radio map.
Although the positioning accuracy can be improved to the order of centimeters using techniques such as angle-of-arrival\cite{travlakis-ojcom2023}, there is generally a trade-off between accuracy and device cost.
The crowdsourcing system depends on user devices, motivating us to develop a radio map construction method that is robust to location errors.
\par
Several works proposed input error-robust regression framework for general regression analysis, not just for radio map construction problems.
For example, \cite{NIPS2011_a8e864d0} introduced a modified Gaussian process regression (GPR) that is resilient to input errors.
This method analyzes input error variance as output error variance using function approximation, enabling tractable yet robust GPR.
However, most of the related methods, including this approach, assume independent and identically distributed (i.i.d.) error in the input information, while straightforward to handle, raises questions about its suitability as an error model in crowdsensing.
Let us consider an environment in which each device collects received power values while moving.
When using GNSS, positioning errors can result from various factors, including multipath propagation caused by obstacles such as buildings and trees, and signal delays or distortions due to ionospheric effects.
This fact suggests that {\it a positioning error might persist in a burst}; in fact, various experimental results in positioning systems report that the estimated positions exhibit bursty errors (e.g., \cite{xiongFaultTolerantGNSSSINS2020}), indicating that error modeling based on the i.i.d. assumption is inadequate in the crowdsourcing-based radio map construction problem.
\par
Based on this background, this study proposes a radio map construction framework which is resilient to the biased location errors.
This method calibrates the location information in the sensing data based on (i)\,distance dependency in the received signal power values and (ii)\,spatial correlation in log-normal shadowing.
To this end, we design a marginal log-likelihood function-based objective function parameterized by location bias factors and statistical information regarding the radio propagation (i.e., path loss index, standard deviation, and spatial correlation).
Maximizing this function estimates the location errors and radio propagation information simultaneously, thereby enabling radio map construction with ex-post-calibrated location information.
Numerical results considering user mobility show that the proposed method improves the accuracy of the radio map construction.

\section{System Model}
\begin{figure}
    \centering
    \includegraphics[width=0.9\linewidth]{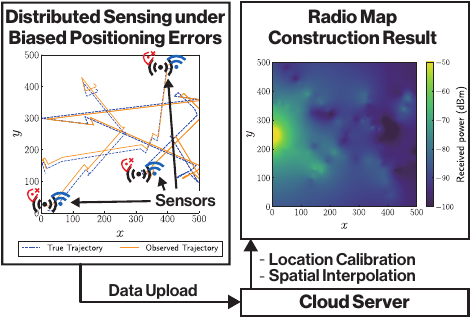}
    \caption{System model.}
    \label{fig:system_model}
\end{figure}
Fig.\,\ref{fig:system_model} illustrates the system model.
This paper considers a radio map construction task in outdoor scenarios; the task is to estimate spatial variation of the received signal power values from a transmitter of interest as accurate as possible based on the on-site sensing and spatial interpolation.
To this end, we consider an environment where $N$ moving sensors distributed in a two-dimensional area observes received signal strength from a transmitter.
The nodes record the received signal power values with the observed positions, and upload them to a server.
The observation positions are acquired by their positioning equipment, such as GNSS; however, the positions are inaccurate due to the sensor biases.
\par
The $i$-th sensor moves between times 1 and $T^{(i)}$, and observes the received signal strength and sensor position every $\tau^{(i)}$ seconds. The total number of data to be observed is thus $M^{(i)} = T^{(i)} / \tau^{(i)}$.
Let $p\left(\mathbf{x}^{(i)}_j\right)$ be the $j$-th observation data by the $i$-th sensor, and let $\mathbf{x}_j^{(i)}$ be its true position.
Each node obtains the received signal power while removing the effects of multipath fading as much as possible to estimate spatial trend of the received signal power values\footnote{Effects of multipath fading in sensing can be mitigated by (i)\, the average of some instantaneous power values obtained in a short time; (ii)\, the use of multiple antennas to take spatial diversity; and (iii)\, the average of signal power values over frequency domain.}.
Then, the received signal strength $p(\mathbf{x})$\,[dBm] observed at $\mathbf{x}$ can be modeled as $p(\mathbf{x}) = f(\mathbf{x}) + \epsilon_p$, where $f(\mathbf{x})$ is the function for received signal power performance, and $\epsilon_p$ is a Gaussian noise caused by uncertainty factors in sensing, including thermal noise, quantization error, and the imperfection in removing the effects of multipath fading.
Considering the path loss and shadowing, $f(\mathbf{x})$ can be expressed as
\begin{align}
    f(\mathbf{x}) &= P_{\mathrm{tx}} - 10 \eta \log_{10}\left(\lVert \mathbf{x}_{\mathrm{tx}} - \mathbf{x} \rVert\right) + w(\mathbf{x})     \label{eq:radio_propagation}\\
    &\triangleq \overline{P}(\mathbf{x}) + w(\mathbf{x}), \label{eq:path_loss}
\end{align}
where $P_\mathrm{tx}$[dBm] is the transmission power, $\eta$ is the path loss index, $\mathbf{x}_{\mathrm{tx}}$ is the transmitter position, and $w(\cdot)$\,[dB] is the shadowing term following a Gaussian process; based on Gudmundson's model\cite{Gudmundson-el1991}, it can be modeled as,
\begin{equation}
    \label{eq:shadowing_gp}
    w(\mathbf{\cdot}) \sim \mathrm{GP}(0, k(\mathbf{x}, \mathbf{x}')),
\end{equation}
where $\mathrm{GP}(\overline{f}(\mathbf{x}), k(\mathbf{x}, \mathbf{x}'))$ is the probability density function (PDF) of GP with mean function $\overline{f}(\mathbf{x})$ and kernel function $k(\mathbf{x}, \mathbf{x}')$. 
According to the Gudmundson's shadowing correlation model, we use the exponential covariance function as the kernel function; this function is given by,
\begin{equation}
    \label{eq:exponential_kernel}
    k(\mathbf{x}, \mathbf{x}') = \sigma_f^2 \exp\left(-\frac{\lVert \left(\mathbf{x} - \mathbf{x}'\right) \rVert \log{(2)}}{d_{\mathrm{cor}}}\right),
\end{equation}
where $\sigma_f$ is the standard deviation of the shadowing term, and $d_{\mathrm{cor}}$ is the correlation length. 
\par
\begin{figure}[t]
    \centering
    \begin{subfigure}[c]{0.45\linewidth}
        \centering
        \includegraphics[width=\linewidth]{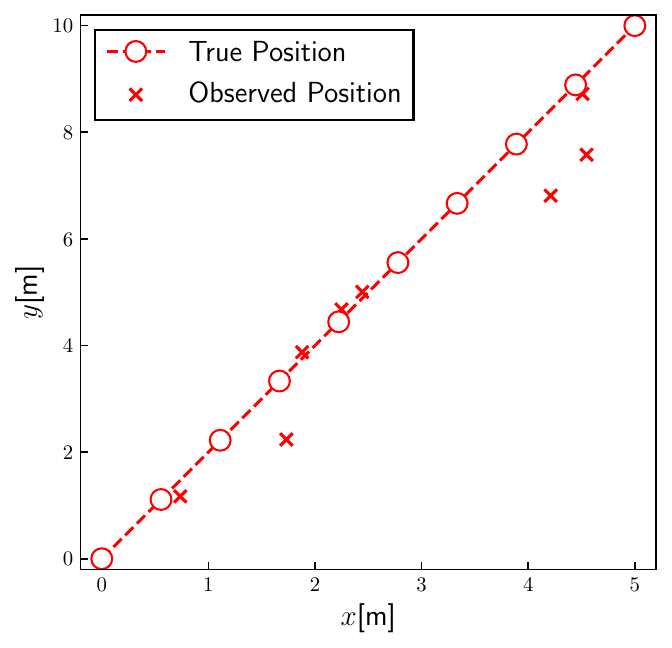}
        \caption{i.i.d. errors.}
    \end{subfigure}
    \begin{subfigure}[c]{0.45\linewidth}
        \centering
        \includegraphics[width=\linewidth]{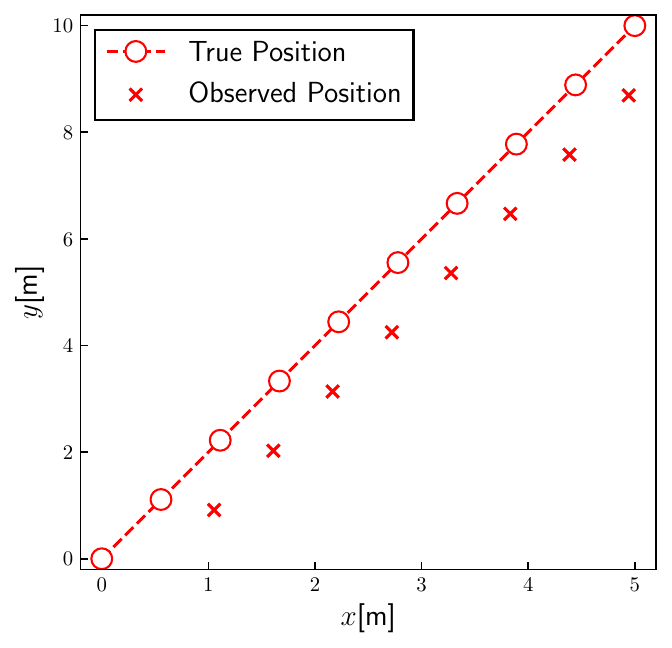}
        \caption{Biased errors.}
    \end{subfigure}
    \caption{Examples of positioning error models.}
    \label{fig:examples_errors}
\end{figure}
Next, we model the positioning error.
GNSS is widely used for a positioning, however, a single positioning operation often takes several tens of seconds under a cold start and a few seconds even with a warm start.
Smartphones can reduce this positioning delay by continuously updating the initial position information using onboard sensors such as accelerometers and Wi-Fi and Bluetooth-based information.
This implementation suggests that the positioning error within a sensing period is {\it not} i.i.d., but is instead expected to exhibit bursty behavior influenced by the initial localization error; even when GNSS continuously observes the position information, it still shows bursty errors due to multipath effects of GNSS signals (e.g.,\,\cite{xiongFaultTolerantGNSSSINS2020}).
To capture this effect, we model the positioning error as a bias error originating from the initial position estimate.
Based on this model, the observed location information $\tilde{\mathbf{x}}_j^{(i)}$ can be expressed as the following equation:
\begin{equation}
    \label{eq:sensor_bias}
    \tilde{\mathbf{x}}_j^{(i)} = \mathbf{x}_j^{(i)} + \boldsymbol{\epsilon}_j^{(i)},
\end{equation}
where $\boldsymbol{\epsilon}_j^{(i)} \sim \mathcal{N}(0, \boldsymbol{\sigma}^{(i)}\mathbf{I})$. Here, $\boldsymbol{\sigma}^{(i)} = \left[\sigma_1^{(i)}, \sigma_2^{(i)}\right]^\top$ is the standard deviation of the sensor bias of the $i$-th sensor and $\mathbf{I}$ is the identity matrix. Note that sensor biases are unique to each sensor and do not vary over $T^{(i)}$.
Examples of positioning errors are summarized in Fig.\,\ref{fig:examples_errors}.
In Fig.\,\ref{fig:examples_errors}(a), the observation points are affected by the i.i.d. Gaussian noise, while Fig.\,\ref{fig:examples_errors}(b) is based on Eq.\,\eqref{eq:sensor_bias}.
Note that while the model assumes a single burst error per sensor, it can support multiple burst errors occurring in rapid succession on a single sensor by separating time-series data into multiple sets.
\par
Finally, the local dataset at the $i$-th sensor can be written as
\begin{equation}
    \mathcal{D}^{(i)} = \left\{\left[\tilde{\mathbf{x}}_j^{(i)}, p\left(\mathbf{x}_j^{(i)}\right)\right] \;\middle|\; j = 1, 2, \cdots, M^{(i)}_j \right\}.
\end{equation}
\par
The server tries to estimate the received signal strength at $\mathbf{x}_\ast$ based on the full dataset $\mathcal{D} = \bigcup_{i=1}^{N} \mathcal{D}^{(i)}$.
Note that we assume major statistical information regarding the received signal power, $\eta$, $d_\mathrm{cor}$, $P_\mathrm{tx}$, are unknown in advance, which will be estimated in the proposed algorithm.
In contrast, the transmission coordinate $\mathbf{x}_\mathrm{tx}$ and standard deviations in location uncertainty, $\boldsymbol{\sigma} = \left[\boldsymbol{\sigma}^{(1)}, \boldsymbol{\sigma}^{(2)}, \cdots, \boldsymbol{\sigma}^{(M)}\right]$, are assumed to be available at the server.

\section{Radio Map Construction Based on Pure GPR}
\begin{algorithm}[t]
    \caption{\strut Radio map construction based on GPR.}
    \label{alg:ordinary_radio_map_construction}
    \begin{algorithmic}[1]
        \State Estimate $(\hat{P}_{\mathrm{tx}},\,\hat{\eta})$ by solving Eq.\,\eqref{eq:ordinary_least_squares}.
        \ForAll{measurement $(\mathbf{x},\,p(\mathbf{x}))$}
            \State Compute $\overline{P}_{\mathrm{est}}(\mathbf{x}) \gets \hat{P}_{\mathrm{tx}} - 10\,\hat{\eta}\,\log_{10}\Bigl\lVert \mathbf{x}_{\mathrm{tx}} - \mathbf{x}\Bigr\rVert$
            \State Compute $\hat{w}(\mathbf{x})$ by Eq.\,\eqref{eq:shadowing_estimation}.
        \EndFor
        \State Estimate $(\hat{\sigma}_f,\,\hat{d}_\mathrm{cor},\,\hat{\sigma}_n)$ by maximizing Eq.\,\eqref{eq:ordinary_likelihood}.
        \State Set the vector $k(\mathbf{x}_\ast,\,\mathbf{X})$ as Eq.\,\eqref{eq:kernel_vector_ast}.
        \State Compute the GP posterior for $\hat{w}(\mathbf{x}_\ast)$ by Eq.\,\eqref{eq:ordinary_shadowing_estimation}.
        \State Compute $p(\mathbf{x}_\ast)$ by Eq.\,\eqref{eq:ordinary_gp_regression}.
        \State Compute $\operatorname{Var}(p(\mathbf{x}_\ast))$ by Eq.\,\eqref{eq:ordinary_gp_regression_variance}.
        \State \Return $p(\mathbf{x}_\ast), \operatorname{Var}p(\mathbf{x}_\ast)$
    \end{algorithmic}
\end{algorithm}

This section introduces the baseline method for the radio map construction based on GPR\cite{Rasmussen2004}, that does not consider the effects of sensor biases\footnote{As mentioned in Sect.\,\ref{sect:introduction}, noisy-input GP\cite{NIPS2011_a8e864d0} (NIGP) can improve the robustness of the regression to the i.i.d. input error. Although it can enhance accuracy in variance computation, its computation results in mean computations are almost equivalent to the pure GPR this section presents; the accuracy improvement shown in Sect.\,\ref{sect:performance} shows a nearly equivalent trend compared to GPR-based radio map construction considering i.i.d. positioning errors. For the sake of simplicity, this paper treats the pure GPR as a baseline.}.
According to Eqs.\,(\ref{eq:radio_propagation}-\ref{eq:shadowing_gp}), the prior distribution regarding the received signal power can be modeled as GP with the path loss-based mean received signal power and the spatially correlated Gaussian distribution; i.e., $f(\mathbf{x}) \sim \mathrm{GP}(\overline{P}(\mathbf{x}), k(\mathbf{x}, \mathbf{x}'))$.
Based on this relationship, GPR estimates the received signal power at $\mathbf{x}_\ast$ so that the posterior distribution on $f(\mathbf{x}_\ast)$ is fitted to the GP constrained by $\mathcal{D}$.
Note that we describe computation method for the received signal power on a target location $\mathbf{x}_\ast$ to sake of simplicity; however, by treating all possible target points within the target area as $\mathbf{x}_\ast$ and applying the proposed method, it is possible to estimate the received power across the entire area.
Major procedures of this method can be categorized into 
the following steps: (i)\,path loss estimation for the prior mean modeling,
 (ii)\,kernel modeling based on maximum likelihood estimation, and (iii)\,prediction the posterior performance regarding $f(\mathbf{x}_\ast)$.
\par
First, this method estimates $\overline{P}(\mathbf{x})$ to form the prior mean in GP.
It requires to estimate the transmission power $P_{\mathrm{tx}}$ and the path loss index $\eta$; we can obtain these information by solving the least squares problem in Eq.\,\eqref{eq:ordinary_least_squares}. Assuming $\hat{P}_{\mathrm{tx}}$ and $\hat{\eta}$ are the estimation results regarding $P_{\mathrm{tx}}$ and $\eta$, respectively, this problem can be written as,
\begin{equation}
    \label{eq:ordinary_least_squares}
    \min_{\hat{P}_{\mathrm{tx}}, \hat{\eta}} \sum_{i=1}^{N} \sum_{j = 1}^{M^{(i)}} \left( p\left(\mathbf{x}_j^{(i)}\right) - \hat{P}_{\mathrm{tx}} - 10 \hat{\eta} \log_{10} \lVert\mathbf{x}_{\mathrm{tx}} - \mathbf{x}_j^{(i)}\rVert \right)^2.
\end{equation}
This paper solves this method based on the ordinary least squares.
Then, the prior mean and the shadowing factor at $\mathbf{x}_j^{(i)}$ can be estimated by the following equations, respectively.
\begin{align}
    \label{eq:shadowing_estimation}
    \overline{P}_{\mathrm{est}}\left(\mathbf{x}_j^{(i)}\right) &\triangleq \hat{P}_\mathrm{tx} - 10 \hat{\eta} \log_{10} \left( \mathbf{x}_{\mathrm{tx}} - \mathbf{x}_j^{(i)} \right),\\
    \hat{w}\left(\mathbf{x}_j^{(i)}\right)
    &= p\left(\mathbf{x}_j^{(i)}\right) - \overline{P}_{\mathrm{est}}(\mathbf{x}).
\end{align}
As shown in Eq.\,\eqref{eq:shadowing_gp}, shadowing factors follow the GP with zero mean and the exponential kernel, allowing us to find $w(\mathbf{x}_\ast)$ by GPR\footnote{Pure GPR incurs a computational cost of $\mathcal{O}\left((M_N)^3\right)$. However, $M_N < 10^4$ in a typical radio map construction problem, and radio maps can be constructed in advance of application operation. Thus, this paper does not address the computational complexity issue.} using an exponential covariance function. \par
The log-likelihood function for the GP is given by
\begin{multline}
    \label{eq:ordinary_likelihood}
    \log p\left(\mathbf{w} \mid \sigma_f, d_{\mathrm{cor}}, \sigma_n \right) = -\frac{1}{2}\mathbf{w}^\top (\mathbf{K} + \sigma_n^2 \mathbf{I})^{-1} \mathbf{w} \\
    - \frac{1}{2}\log\left(\lvert \mathbf{K} + \sigma_n^2 \mathbf{I} \rvert\right) - \frac{M_N}{2}\log(2\pi),
\end{multline}
where $M_N = \sum_{i=1}^N M^{(i)}$, $\mathbf{K}$ is the $M_N \times M_N$ covariance matrix, $\sigma_n$ is the noise standard deviation, $\mathbf{I}$ is the $M_N \times M_N$ identity matrix, and
\begin{equation}
    \mathbf{w} \!=
    \!\!\left[ \hat{w}\left(\mathbf{x}_1^{(1)}\!\right), \hat{w}\left(\mathbf{x}_2^{(1)}\!\right), \cdots, \hat{w}\left(\mathbf{x}_{M_1}^{(1)}\!\right), \cdots, \hat{w}\left(\mathbf{x}_{M_N}^{(N)}\!\right) \!\right]^\top.\nonumber
\end{equation}
The parameters $\sigma_f$, $d_\mathrm{cor}$, and $\sigma_n$ can be estimated by maximizing the log-likelihood function given in Eq.\,\eqref{eq:ordinary_likelihood}, based on a gradient-based or black-box optimization algorithm. The corresponding estimates are denoted by $\hat{\sigma}_f$, $\hat{d}_{\mathrm{cor}}$ and $\hat{\sigma}_n$.
After the parameters are estimated, the estimated value of the shadowing factor at $\mathbf{x}_\ast$ and its variance can be expressed as
\begin{align}
    \hat{w}(\mathbf{x}_\ast) &= k(\mathbf{x}_\ast, \mathbf{X})^\top \left(\mathbf{K} + \hat{\sigma}_n^2 \mathbf{I}\right)^{-1} \mathbf{w},    \label{eq:ordinary_shadowing_estimation}\\
    \operatorname{Var}\left(\hat{w}(\mathbf{x}_\ast)\right) &= k(\mathbf{x}_\ast, \mathbf{x}_\ast) \nonumber \\
    &- k(\mathbf{x}_\ast, \mathbf{X})^\top \left(\mathbf{K} + \hat{\sigma}_n^2 \mathbf{I}\right)^{-1} k(\mathbf{x}_\ast, \mathbf{X}),\label{eq:ordinary_shadowing_variance}
\end{align}
where
\begin{align}
    k(\mathbf{x}_\ast, \mathbf{X}) &= \left[k\left(\mathbf{x}_\ast, \mathbf{x}_1^{(1)}\right), \cdots, k\left(\mathbf{x}_\ast, \mathbf{x}_{M_N}^{(N)}\right)\right]^\top.\label{eq:kernel_vector_ast}
\end{align}
Finally, the received signal strength $p(\mathbf{x}_\ast)$ and variance of $p(\mathbf{x}_\ast)$ at can be estimated by
\begin{align}
    p(\mathbf{x}_\ast) &= \overline{P}_{\mathrm{est}}(\mathbf{x}_\ast) + \hat{w}(\mathbf{x}_\ast), \label{eq:ordinary_gp_regression}\\
    \operatorname{Var}\left(p(\mathbf{x}_\ast)\right) &= \operatorname{Var}\left(\hat{w}(\mathbf{x}_\ast)\right).\label{eq:ordinary_gp_regression_variance}
\end{align}
This algorithm is summarized in Algorithm\,\ref{alg:ordinary_radio_map_construction}.

\section{Joint Location Calibration and\\Radio Map Construction}
\label{ssec:proposed_radio_environment_map_construction}
Since Alg.\,\ref{alg:ordinary_radio_map_construction} does not consider any location uncertainty, its prediction accuracy would be degraded if the location noise is significant.
Aiming to construct an accurate radio map in the noisy-location condition, we propose a joint location calibration and radio map construction.
In particular, leveraging the fact that the received power follows a GP, we design a likelihood function that accounts for the location bias factors.
Based on inference via maximization of this log-likelihood-based objective function, our method jointly estimates the set of location biases and the parameters in received signal power performance, thereby enabling us to construct an accuracy-enhanced radio map with the calibrated location information.
\par
We integrate the location bias factors into the radio map construction problem by focusing on the radio propagation characteristics.
In particular, we propose estimating path loss and shadowing factor together as in Eq.\,\eqref{eq:proposed_radio_propagation_gp}, so that the model considers both location biases and radio propagation characteristics simultaneously and avoids overfitting the location biases to either the path loss or shadowing;
\begin{equation}
    \label{eq:proposed_radio_propagation_gp}
    f(\mathbf{x}) \sim \mathrm{GP}(\mu(\mathbf{x}), k_{\mathrm{PL}}(\mathbf{x}, \mathbf{x}') + k_{\mathrm{shad}}(\mathbf{x}, \mathbf{x}')),
\end{equation}
where $\mu(\mathbf{x})$ is the mean function, $k_{\mathrm{PL}}(\mathbf{x}, \mathbf{x}')$ is a kernel function corresponding to the path loss factor, and $k_{\mathrm{shad}}(\mathbf{x}, \mathbf{x}')$ is a kernel function corresponding to the shadowing factor.
Since both path loss and shadowing factors are incorporated into the kernel function, the mean function $\mu(\mathbf{x})$ can be considered a constant function $\mu(\mathbf{x}) = \theta_1$, where $\theta_1 \in \mathbb{R}$.
According to Eq.\,\eqref{eq:path_loss}, The path loss factor is logarithmic decay with respect to the distance between the transmitter and the receiver.
$k_{\mathrm{PL}}(\mathbf{x}, \mathbf{x}')$ can be written as in Eq.\,\eqref{eq:path_loss_kernel}
\begin{multline}
    \label{eq:path_loss_kernel}
    k_{\mathrm{PL}}(\mathbf{x}, \mathbf{x}') = \\
    \theta_2 \exp\left(-\frac{\left|\log_{10}{\lVert\mathbf{x}_{\mathrm{tx}} - \mathbf{x}\rVert} - \log_{10}{\lVert\mathbf{x}_{\mathrm{tx}} - \mathbf{x}'\rVert}\right|^2}{2\theta_3}\right),
\end{multline}
where $\theta_2$ and $\theta_3$ are positive-valued hyperparameters. This kernel function was constructed to express the shape of an RBF kernel, based on the characteristics of the path loss.
According to Eqs.\,(\ref{eq:shadowing_gp},\,\ref{eq:exponential_kernel}), $k_{\mathrm{shad}}(\mathbf{x}, \mathbf{x}')$ is $k(\mathbf{x}, \mathbf{x}')$ with the $\sigma_f$ and $d_{\mathrm{cor}}$ parameters replaced by $\theta_4$ and $\theta_5$, respectively.
Further, from Eq.\,\eqref{eq:sensor_bias}, the actual position can be expressed as $\mathbf{x}_j^{(i)} = \tilde{\mathbf{x}}_j^{(i)} - \boldsymbol{\epsilon}^{(i)}$, suggesting that $k_{\mathrm{PL}}(\mathbf{x}, \mathbf{x}')$ and $k_{\mathrm{shad}}(\mathbf{x}, \mathbf{x}')$ can be rewritten as
\begin{IEEEeqnarray}{lr}
    \label{eq:path_loss_kernel_bias}
    k_{\mathrm{PL}}\left(\tilde{\mathbf{x}}_j^{(i)}, \tilde{\mathbf{x}}_k^{(l)}\right) = \\
    \theta_2 \exp\left(-\frac{\left|\log_{10}{\left\lVert\mathbf{x}_{\mathrm{tx}} - d_j^{(i)} \right\rVert} - \log_{10}{\left\lVert\mathbf{x}_{\mathrm{tx}} - d_k^{(l)} \right\rVert}\right|^2}{2\theta_3}\right), \nonumber
\end{IEEEeqnarray}
\begin{equation}
    \label{eq:shadowing_kernel_bias}
    k_{\mathrm{shad}}\left(\tilde{\mathbf{x}}_j^{(i)}, \tilde{\mathbf{x}}_k^{(l)}\right) = \theta_4 \exp\left(-\frac{\lVert d_j^{(i)} - d_k^{(l)} \rVert}{2\theta_5}\right),
\end{equation}
where
\begin{align}
        d_j^{(i)} &\triangleq \left(\tilde{\mathbf{x}}_j^{(i)} - \hat{\boldsymbol{\epsilon}}^{(i)}\right), \\
        d_k^{(l)} &\triangleq \left(\tilde{\mathbf{x}}_k^{(l)} - \hat{\boldsymbol{\epsilon}}^{(j)}\right),
\end{align}
and
\begin{equation}
    \hat{\boldsymbol{\epsilon}}^{(i)} = \left[ \epsilon_1^{(i)}, \epsilon_2^{(i)} \right]^\top
\end{equation}
is the hyperparameter vector, and corresponds to the estimated bias of the $i$-th sensor. \par
Since the probability density function for GP can be derived by a multivariate Gaussian distribution, the marginal log-likelihood function with this kernel design is given by
\begin{multline}
    \label{eq:proposed_likelihood}
    \log \mathcal{L}(\mathbf{p} \mid \boldsymbol{\theta}, \hat{\boldsymbol{\epsilon}}) = - \frac{M_N}{2}\log(2\pi)\\
    -\frac{1}{2}(\mathbf{p} - \theta_1)^\top (\mathbf{K}_{\mathrm{PL}} + \mathbf{K}_{\mathrm{shad}} + \theta_6 \mathbf{I})^{-1} (\mathbf{p} - \theta_1) \\
    - \frac{1}{2}\log\left(\lvert \mathbf{K}_{\mathrm{PL}} + \mathbf{K}_{\mathrm{shad}} + \theta_6 \mathbf{I} \rvert\right),
\end{multline}
where $\boldsymbol{\theta} = \left[\theta_1, \theta_2, \theta_3, \theta_4, \theta_5, \theta_6\right]$, $\mathbf{K}_{\mathrm{PL}}$ and $\mathbf{K}_{\mathrm{shad}}$ are kernel matrices whose each element is given by Eqs.\,(\ref{eq:path_loss_kernel_bias},\,\ref{eq:shadowing_kernel_bias}), and both have a size of $M_N \times M_N$.
Further, $\mathbf{p}$ and $\hat{\boldsymbol{\epsilon}}$ are the following vectors, respectively.
\begin{align}
    \mathbf{p} &= \left[p_1^{(1)}, p_2^{(1)}, \cdots, p_{M_1}^{(1)}, \cdots, p_{M_N}^{(N)}\right]^\top, \\
    \hat{\boldsymbol{\epsilon}} &= \left[\hat{\boldsymbol{\epsilon}}^{(1)}, \hat{\boldsymbol{\epsilon}}^{(2)}, \cdots, \hat{\boldsymbol{\epsilon}}^{(N)}\right],
\end{align}
\par
Typically, maximizing the likelihood function can find the parameters of interest.
However, this problem can show multiple solutions for the hyperparameters $\hat{\boldsymbol{\epsilon}}$.
This is because when a certain value of $\hat{\boldsymbol{\epsilon}}'$ is a solution, $\hat{\boldsymbol{\epsilon}}' + \mathbf{r}$ can also be a solution for any $\mathbf{r} \in \mathbb{R}^2$.
To solve this problem, we introduce a penalty factor based on the standard deviation of sensor bias $\boldsymbol{\sigma}$.
For the penalty factor, let us consider the probability density of a two-dimensional independent multivariate normal distribution with mean 0 and standard deviation $\boldsymbol{\sigma}^{(i)}$, which is given by,
\begin{multline}
    \label{eq:independent_multivariate_normal}
    \phi(\mathbf{x}; 0, \boldsymbol{\sigma}^{(i)}) = \\
    \frac{1}{2\pi \sigma_1^{(i)} \sigma_2^{(i)}} \exp\left(-\frac{1}{2} \left( \frac{x_1^2}{\left(\sigma_1^{(i)}\right)^2} + \frac{x_2^2}{\left(\sigma_2^{(i)}\right)^2} \right) \right),
\end{multline}
where $\mathbf{x} = \left[x_1, x_2\right]^\top$.
By adding $\sum_{i=1}^{N} \log \phi(\mathbf{x}; 0, \boldsymbol{\sigma}^{(i)})$ to the marginal log-likelihood, we have
\begin{IEEEeqnarray}{lr}
    \label{eq:proposed_likelihood_sigma}
    \log \mathcal{L}'(\mathbf{p} \mid \boldsymbol{\theta}, \hat{\boldsymbol{\epsilon}}, \boldsymbol{\sigma}) = \log \mathcal{L}(\mathbf{p} \mid \boldsymbol{\theta}, \hat{\boldsymbol{\epsilon}}) \\
    - \sum_{i=1}^{N}\left\{\log\left( 2\pi \sigma_1^{(i)} \sigma_2^{(i)} \right) + \frac{1}{2} \left( \frac{\left(\hat{\epsilon}^{(i)}_1\right)^2}{\left(\sigma_1^{(i)}\right)^2} + \frac{\left(\hat{\epsilon}^{(i)}_2\right)^2}{\left(\sigma_2^{(i)}\right)^2} \right)\right\}. \nonumber
\end{IEEEeqnarray}
Maximizing Eq.\,\eqref{eq:proposed_likelihood_sigma} based on a gradient-based optimization can find the optimal value of $\hat{\boldsymbol{\epsilon}}$, denoted as $\hat{\boldsymbol{\epsilon}}_{\mathrm{opt}}$.
Once the bias factors are estimated, the location information can be then corrected to $\hat{\mathbf{x}}_j^{(i)} = \tilde{\mathbf{x}}_j^{(i)} - \hat{\boldsymbol{\epsilon}}_{\mathrm{opt}}^{(i)}$, where $\hat{\mathbf{x}}_j^{(i)}$ is the corrected sensor position.
Finally, we can obtain the accuracy-enhanced radio map by Alg.\,\ref{alg:ordinary_radio_map_construction} and $\hat{\mathbf{x}}_j^{(i)}$.

\section{Performance Evaluation}
\label{sect:performance}
This section evaluates the accuracy difference between radio maps constructed using sensor positions corrected by the proposed method and those built without any sensor position correction. We used gradient descent with Schedule-Free\,\cite{defazioRoadLessScheduled2024} and RAdam\,\cite{liuVarianceAdaptiveLearning2020} for bias estimation, and Nelder-Mead\,\cite{gaoImplementingNelderMeadSimplex2012} for shadowing estimation. \par
\begin{figure}[t]
    \centering
    \includegraphics[width=1.0\linewidth]{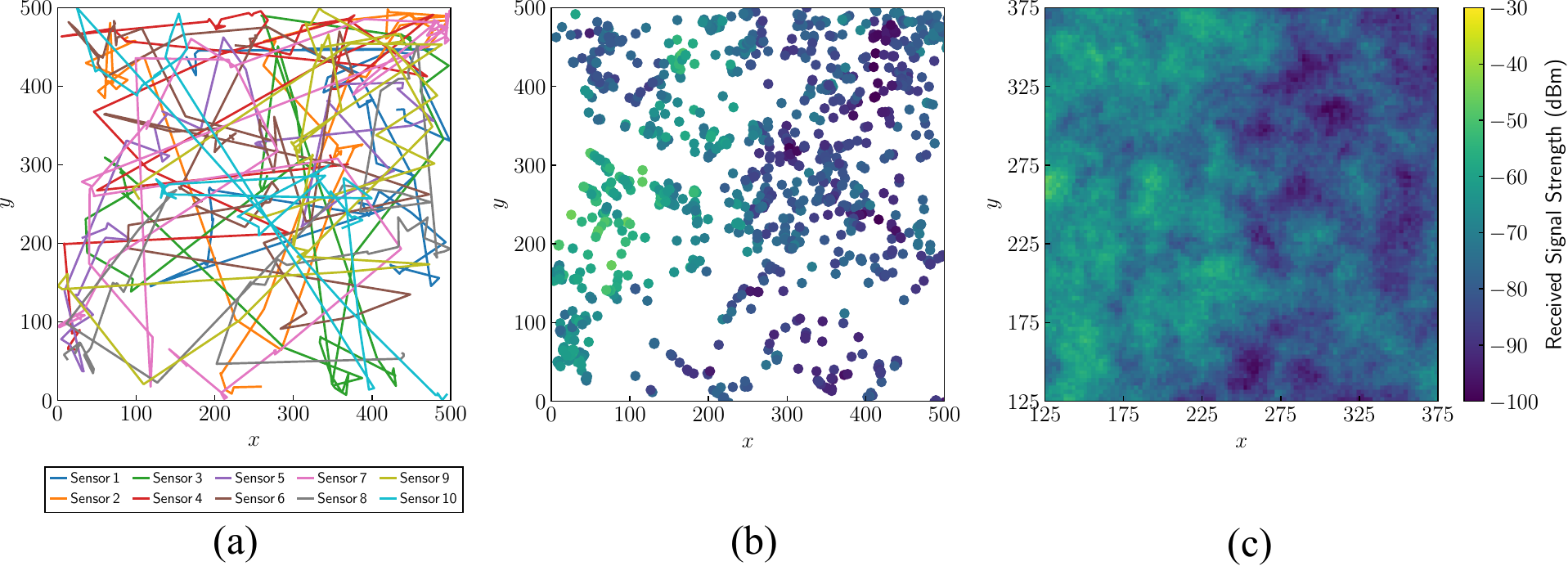}
    \caption{Data example. (a)\, sensor trajectories, (b)\,observation data, and (c)\,ground-truth radio map.}
    \label{fig:sensor_movement_and_rss}
\end{figure}

The sensor movement was assumed to follow the Levy-Walk\,\cite{rheeLevyWalkNatureHuman2011}, which mimics human movement, and Levy-Walk parameters $\alpha$ and $\beta$ were set to 0.5 and 1.0, respectively.
The sensor's moving space was set to be $500\mathrm{m} \times 500\mathrm{m}$. The transmitter was placed at $[0, 250]$, the transmission power was set to 10 dBm, and the path loss index $\eta$ was set to 4.0. The shadowing factor was modeled as a Gaussian process with a standard deviation of 8.0 dB and a correlation length of 20.0 m.
The number of sensors $N$ was set to 10, and the standard deviation of the sensor bias $\boldsymbol{\sigma}$ was set to $[10 [\mathrm{m}], 10 [\mathrm{m}]]^\top$.
Each evaluation was carried out according to the conditions shown in Table\,\ref{tab:evaluation_condition}, with each setting evaluated over 1000 independent trials; note that Fig.\,\ref{fig:sensor_movement_and_rss} shows an example of the dataset under Exp 1 including (a)\,sensor trajectories, (b)\,observation data, and (c)\,ground-truth radio map.
\begin{table}[!t]
    \centering
    \caption{Evaluation conditions.}
    \label{tab:evaluation_condition}
    \begin{tabular}{c|c|c|c|c} \hline
        \textbf{Parameter} & \textbf{Exp\,1} & \textbf{Exp\,2} & \textbf{Exp\,3} & \textbf{Exp\,4} \\ \hline
        $T$[s] & 3600 & 7200 & 1800 & 900 \\ 
        $\tau$[s] & 20 & 40 & 10 & 5 \\ 
        \hline
    \end{tabular}
\end{table}
The accuracy of both the radio map construction and the bias estimation was evaluated by root mean square error (RMSE). In particular, the accuracy of the radio map construction was assessed within the central $250\mathrm{[m]} \times 250\mathrm{[m]}$ region of the sensors' movement area. \par
\begin{figure}[t]
    \centering
    \includegraphics[width=\linewidth]{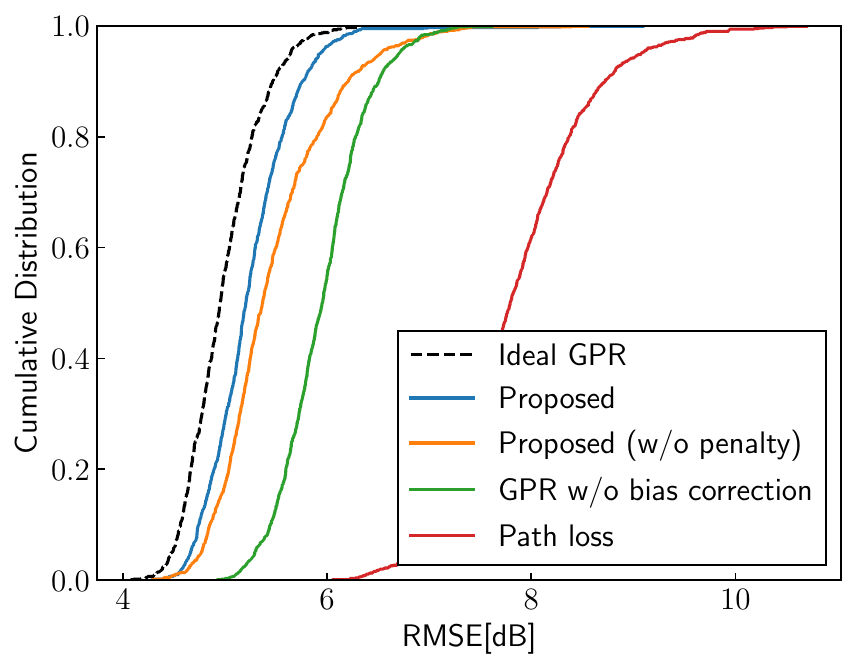}
    \caption{RMSE performance in radio map construction.}
    \label{fig:radio_map_accuracy}
\end{figure}
Fig.\,\ref{fig:radio_map_accuracy} shows the cumulative distribution of RMSE in radio map construction in Exp\,1.
``Ideal GPR'' represents the RMSE when the radio map is constructed using the noise-free sensor positions, indicating an ideal performance in this simulation.
``Proposed'' shows the RMSE obtained by the proposed method, which includes a regularization term as defined in Eq.\,\eqref{eq:proposed_likelihood_sigma}, while ``Proposed (w/o penalty)'' denotes the RMSE by the proposed method but without the regularization term, i.e., based on Eq.\,\eqref{eq:proposed_likelihood}; we evaluated this method to verify effects of the regularization term.
Further, ``GPR w/o bias correction'' corresponds to the RMSE when the radio map is constructed without applying any bias estimation or correction.
Finally, ``Path loss'' indicates the RMSE when the radio map is constructed solely based on path loss estimation, without employing GPR or any other advanced modeling techniques. \par
\begin{figure}[t]
    \centering
    \includegraphics[width=\linewidth]{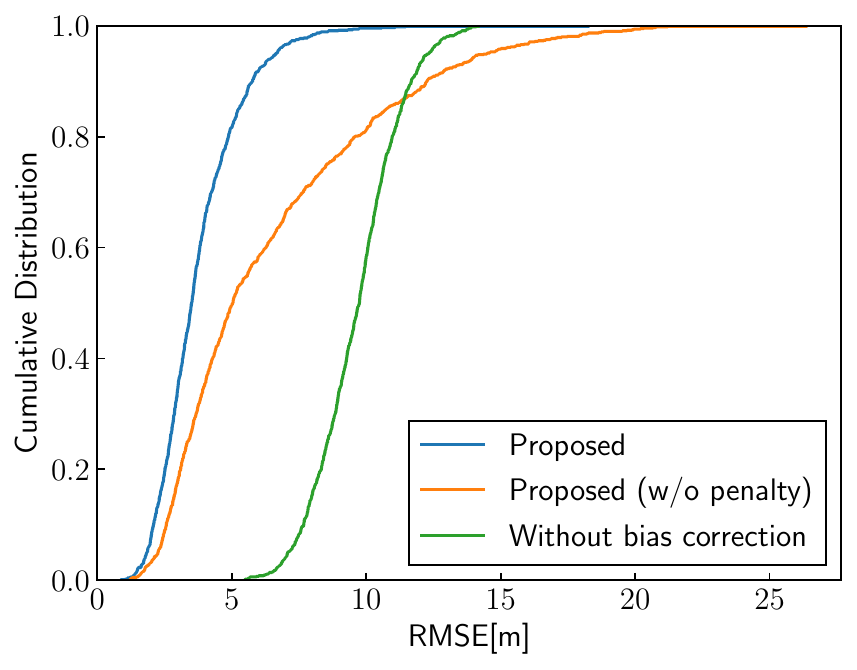}
    \caption{RMSE performance in bias estimation.}
    \label{fig:bias_estimation_accuracy}
\end{figure}
Fig.\,\ref{fig:bias_estimation_accuracy} also shows the cumulative distribution of RMSE in the bias estimation in Exp\,1.
In addition to the proposed method-based designs, this figure also shows the performance of ``Without bias correction'', which corresponds to the RMSE when the bias is not corrected.
The results of the radio map construction accuracy are shown in Table.\,\ref{tab:RadioMap_percentile} and Table\,\ref{tab:Bias_percentile}. \par
Compared with the ideal GPR, the proposed method shows a slight performance degradation of approximately 0.25 dB (50 percentile) and 0.29 dB (90 percentile).
In contrast, the GPR without bias correction exhibits a more substantial degradation of about 1.00 dB (50 percentile) and 1.04 dB (90 percentile), while using path loss significantly worsens the performance by approximately 
2.84 dB (50 percentile) and 3.25 dB (90 percentile).
Thus, the proposed method effectively reduces effects of bias errors.
Similar trends were consistently observed in all experiments (Exp\,1 to Exp\,4), where the proposed method maintained performance degradation close to ideal GPR and outperformed the comparison methods.
The proposed method also achieved high accuracy in bias estimation, improving accuracy by approximately 6.24 m (50 percentile) and 5.90 m (90 percentile) compared to the method without bias estimation.
Notably, the performance of the proposed method without the penalty term deteriorates significantly, with the RMSE in radio map construction worsening by approximately 0.41 dB (50 percentile) and 0.74 dB (90 percentile) compared to the proposed method with the penalty term. This is also the case with bias estimation.
Additionally, the performance difference between the proposed methods with and without the penalty term becomes more significant as the observation time $T$ decreases, demonstrating the effects of the regularization.
\begin{table}[t]
    \centering
    \caption{RMSE in radio map construction. The most accurate methods under location-error conditions are \textbf{bolded}.}
    \label{tab:RadioMap_percentile}
    \begin{subtable}{\linewidth}
        \caption{50 percentile (median)}
        \centering
        \begingroup
        \setlength{\tabcolsep}{2pt}
        \footnotesize
            \begin{tabular}{c|c:cccc} \hline \hline
                & \multirow{2}{*}{Ideal GPR} & \multirow{2}{*}{Proposed} & Proposed & GPR w/o & \multirow{2}{*}{Path loss} \\
                & &  & (w/o penalty) & bias correction &  \\ \hline
                Exp\,1 & 4.95 & \textbf{5.20} & 5.37 & 5.95 & 7.79 \\
                Exp\,2 & 4.71 & \textbf{4.95} & 4.97 & 5.81 & 7.83 \\
                Exp\,3 & 5.23 & \textbf{5.56} & 6.06 & 6.09 & 7.89 \\
                Exp\,4 & 5.53 & \textbf{5.91} & 6.77 & 6.28 & 7.86 \\ \hline \hline
            \end{tabular}
        \endgroup
    \end{subtable}
    \begin{subtable}{\linewidth}
        \caption{90 percentile.}
        \centering
        \begingroup
        \setlength{\tabcolsep}{2pt}
        \footnotesize
            \begin{tabular}{c|c:cccc} \hline \hline
                & \multirow{2}{*}{Ideal GPR} & \multirow{2}{*}{Proposed} & Proposed & GPR w/o & \multirow{2}{*}{Path loss} \\
                & &  & (w/o penalty) & bias correction &  \\ \hline
                Exp\,1 & 5.47 & \textbf{5.76} & 6.21 & 6.51 & 8.72 \\
                Exp\,2 & 5.11 & \textbf{5.40} & 5.45 & 6.34 & 8.80 \\
                Exp\,3 & 5.88 & \textbf{6.17} & 7.21 & 6.70 & 8.93 \\
                Exp\,4 & 6.28 & \textbf{6.70} & 7.80 & 6.99 & 8.94 \\ \hline \hline
            \end{tabular}
        \endgroup
    \end{subtable}
\end{table}
\section{Conclusion}
We proposed a radio map construction method that addresses biased positioning errors by simultaneously estimating radio propagation characteristics and bias factors.
Numerical results considering actual human mobility show that the our method outperforms baseline in both radio map and bias estimation accuracy; for example, it limits RMSE degradation to about 0.25-0.29 dB compared to the ideal GPR, while baseline methods exhibit losses over 1 dB.
Our method can enhance the performance of radio map applications, path planning, and resource allocation, where accurate radio maps are essential.
\par
In the future, we will extend the target of the noise model to not only burst errors, but also other types of positioning errors, such as i.i.d. errors.
\begin{table}[t]
    \centering
    \caption{RMSE in bias estimation.}
    \label{tab:Bias_percentile}
    \begin{subtable}{\linewidth}
        \caption{50 percentile (median).}
        \centering
        \begingroup
        \setlength{\tabcolsep}{2pt}
        \footnotesize
            \begin{tabular}{c|ccc} \hline \hline
                & \multirow{2}{*}{Proposed} & Proposed & Without \\
                & & (w/o penalty) & bias correction \\ \hline
                Exp\,1 & \textbf{3.51} & 5.04 & 9.75 \\
                Exp\,2 & \textbf{2.81} & 3.08 & 9.75 \\
                Exp\,3 & \textbf{4.74} & 11.75 & 9.75 \\
                Exp\,4 & \textbf{6.53} & 18.46 & 9.75 \\ \hline \hline
            \end{tabular}
        \endgroup
    \end{subtable}
    \begin{subtable}{\linewidth}
        \caption{90 percentile.}
        \centering
        \begingroup
        \setlength{\tabcolsep}{2pt}
        \footnotesize
            \begin{tabular}{c|ccc} \hline \hline
                & \multirow{2}{*}{Proposed} & Proposed & Without \\
                & & (w/o penalty) & bias correction \\ \hline
                Exp\,1 & \textbf{5.77} & 12.26 & 11.68 \\
                Exp\,2 & \textbf{4.68} & 5.55 & 11.68 \\
                Exp\,3 & \textbf{7.45} & 20.28 & 11.68 \\
                Exp\,4 & \textbf{9.79} & 25.85 & 11.68 \\ \hline \hline
            \end{tabular}
        \endgroup
    \end{subtable}
\end{table}


\end{document}